# Prototype design of a digital Low-Level RF system for $S^3$FEL S-band Transverse Deflecting Cavities

**Jinfu Zhu[a*], Hongli Ding[b†], Haokui Li[b], Jiahang Shao[a], Yong Yu[a], Zongbin Li[a], Jiayue Yang[b], Zhichao Chen[b], Guorong Wu[b] and Weiqing Zhang[b]**

[a] *Institute of Advanced Science Facilities,*
*Xinhu Street, Zhenyuan Road No. 268, Shenzhen, China*
[b] *Dalian Institute of Chemical Physics, Chinese Academy of Sciences,*
*Zhongshan Road No. 457, Dalian, China*

*E-mail*: zhujinfu@mail.iasf.ac.cn, dinghongli@dicp.ac.cn

ABSTRACT: Transverse Deflecting Cavities (TDCs) are generally adopted for electron beam diagnosis. Three sets of S-band and two sets of X-band TDCs are planned at Shenzhen Superconducting Soft X-ray Free Electron Laser ($S^3$FEL) to accurately measure the temporal distribution of ultra-short electron bunches. The microwave system of one TDC consisting of a Low-Level Radio-Frequency system (LLRF), a solid-state amplifier, a klystron, and several waveguide couplers is operated in pulse mode with a maximum repetition rate of 50 Hz. Its microwave stabilities for amplitude and phase are required to be better than 0.05%/0.05° (RMS). This article will introduce the prototype design of the hardware, firmware, and software of the digital LLRF system for S-band TDCs. We use a homemade local oscillator and commercial cards based on the MicroTCA standard in hardware design. The firmware design will use an IQ demodulation and a reference-tracking algorithm to eliminate the measurement noise and drift. The software design is based on the Experimental Physics and Industrial Control System (EPICS), achieving data acquisition, slow control, and interface display functions. This technical report will also show some preliminary test results.

* Corresponding author.
† Corresponding author.

## Contents

## 1. Introduction

As these continuous-wave superconducting facilities of LCLS-II [1], Eu-XFEL [2], and SHINE [3], Shenzhen Superconducting Soft X-ray Free Electron Laser ($S^3$FEL) [4] is also designed to produce high repetition rate X-ray free-electron laser (FEL) pulses of up to megahertz (MHz). The FEL pulses hold the characteristics of ultra-short duration,  high spatial coherence, high peak brightness, and average brightness, providing an exceptional tool for many cutting-edge research fields such as physical chemistry, materials, biological science, and so on [5], [6].

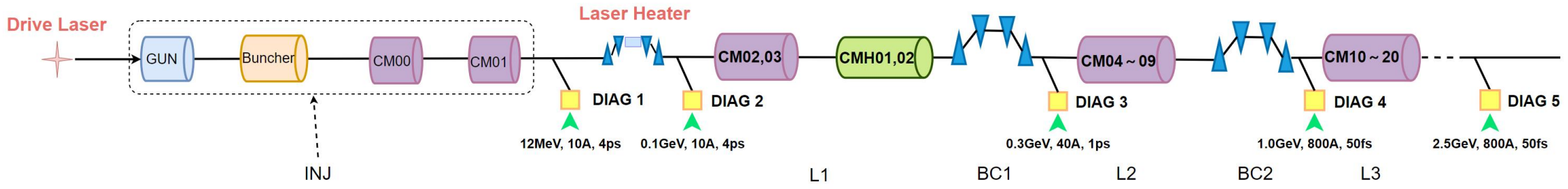

**Figure 1.** The preliminary schematic diagram of the $S^3$FEL LINAC.

Figure 1 shows the preliminary schematic diagram of the $S^3$FEL linear accelerator (LINAC). Located in a 760-meter-long tunnel, the LINAC produces the electron beam with an energy of 2.5 GeV and a repetition rate of 1 MHz. It primarily consists of an injector (INJ), a laser heater, three main linear accelerating sections (L1~L3), two magnetic bunch compressors (BC1~BC2), and diagnostic lines (DIAG1~DIAG5). The injector facilitates beam generation and pre-acceleration, incorporating a driving laser [7], a VHF electron gun [8], an L-band continuous-wave buncher [9], and two 1.3 GHz superconducting cryomodules (CM00~CM01) [10]. The beam energy, peak current, and bunch length at the injector exit are 0.1 GeV, 10 A, and 4 ps, respectively. The main accelerating sections L1 (CM02~CM03, CMH01~CMH02), L2 (CM04~CM09), and L3 (CM10~CM20) boost the beam energy from 0.1 GeV to 2.5 GeV, where CMH represents the 3.9 GHz RF superconducting cryomodule [11]. Additionally, two bunch compressors, BC1/BC2, are employed to compress the bunch length to 1 ps and 50 fs, respectively, and the corresponding beam peak current is increased to 40 A and 800 A. Five beam diagnostic lines are set following the injector, laser heater, BC1, BC2, and beam distribution's output to diagnose these longitudinal beam parameters. Three sets of S-band (2997 MHz) Transverse Deflecting Cavities (TDCs) for DIAG1~DIAG3 and two set of X-band (11989 MHz) TDCs for DIAG4~DIAG5 are deployed to characterize the ultra-short beam temporal distribution accurately [12]. Moreover, several dipole magnets are also used to measure the beam energy distribution. The beam longitudinal phase space can be reconstructed at each diagnostic line by integrating the TDC and dipole magnet. As a result, the longitudinal key parameters of the electron beam can be determined, such as the beam energy, the bunch length, the peak current, and so on.

Due to the limitations imposed by room temperature structure and klystron technology, microwave TDCs cannot operate at a repetition rate of 1 MHz at megawatt level power [13], [14]. Instead, they generally operate at a lower repetition rate of less than 1 kHz, and provided detailed characterization of the electron beam [15]-[18]. Low-Level RF (LLRF) systems provide precise and controllable microwave power in terms of amplitude and phase, drives the operation of the TDC, and maintains stable electromagnetic fields inside the cavity. In recent years, significant progress has been made in developing LLRF systems [19]-[23], particularly in applying digital signal processing technology, such as IQ demodulation, digital filtering, and adaptive control algorithms, greatly enhance the flexibility, accuracy, and real-time feedback capability. These technologies enable LLRF systems to more accurately capture and process the time structure information of electron beams, thereby achieving more efficient time structure diagnosis.

Figure 1 illustrates five sets of TDCs. When the electron beam energy is boosted to 1 GeV and the bunch length is compressed to 50 fs after BC2, the S-band TDC struggles to meet the longitudinal resolution requirements, necessitating a higher-resolution X-band TDC. Although the RF bands are distinguished, the basic structure of the microwave system for the X-band remains consistent with that of the S-band. Therefore, the following introduction will concentrate on the S-band microwave system.

The microwave system mainly consists of a LLRF system, a Solid-State Amplifier (SSA), and a klystron, operating in pulse mode with a maximum repetition rate of 50 Hz. According to electron beam dynamics requirements of the $S^3$FEL LINAC, microwave amplitude stability and phase stability are required to be better than 0.05% (RMS) and 0.05° (RMS), respectively. These S-band stability requirements are related to the temporal resolution of TDC, which is determined as one-twentieth of the RMS bunch length at DIAG3 (i.e., one ps), namely 50 fs. This value corresponds to ~0.05° phase stability at 2997 MHz RF frequency. It is also equivalent to ~0.05% RF field amplitude stability, as the deflecting force is proportional to the product of the field amplitude $V_{\perp}$ and phase $\varphi_{rf}$ near the TDC zero-crossing phase [24].

This technical report will be structured as follows to introduce the prototype design of the digital LLRF for S-band TDCs. Section 2 will show the hardware design of the LLRF system. Section 3 will introduce the firmware and software design, and section 4 will present experiments and results.

## 2. Hardware design

The block diagram of the LLRF system and its connected external systems is depicted in figure 2. The LLRF hardware includes a homemade local oscillator (LO), commercial RTM boards (DWC8VM1) [25], and AMC boards (SIS8300-KU) [26] based on the MicroTCA (MTCA) standard [27].

The optical synchronization system [28] generates RF references for $S^3$FEL TDCs. The master clock signal of the synchronization system is from an RF generator with less than 10 fs [10 Hz, 10 MHz] jitter. It is drift-freely transferred via a stabilization unit of pulsed laser fiber link. 1.3 GHz signal is regenerated by a balanced optical microwave phase detector unit (BOMPD-VCO) [29], [30] based on a pulsed laser train. Then, a frequency synthesizer converts 1.3 GHz to 2997 MHz and 11989 MHz.

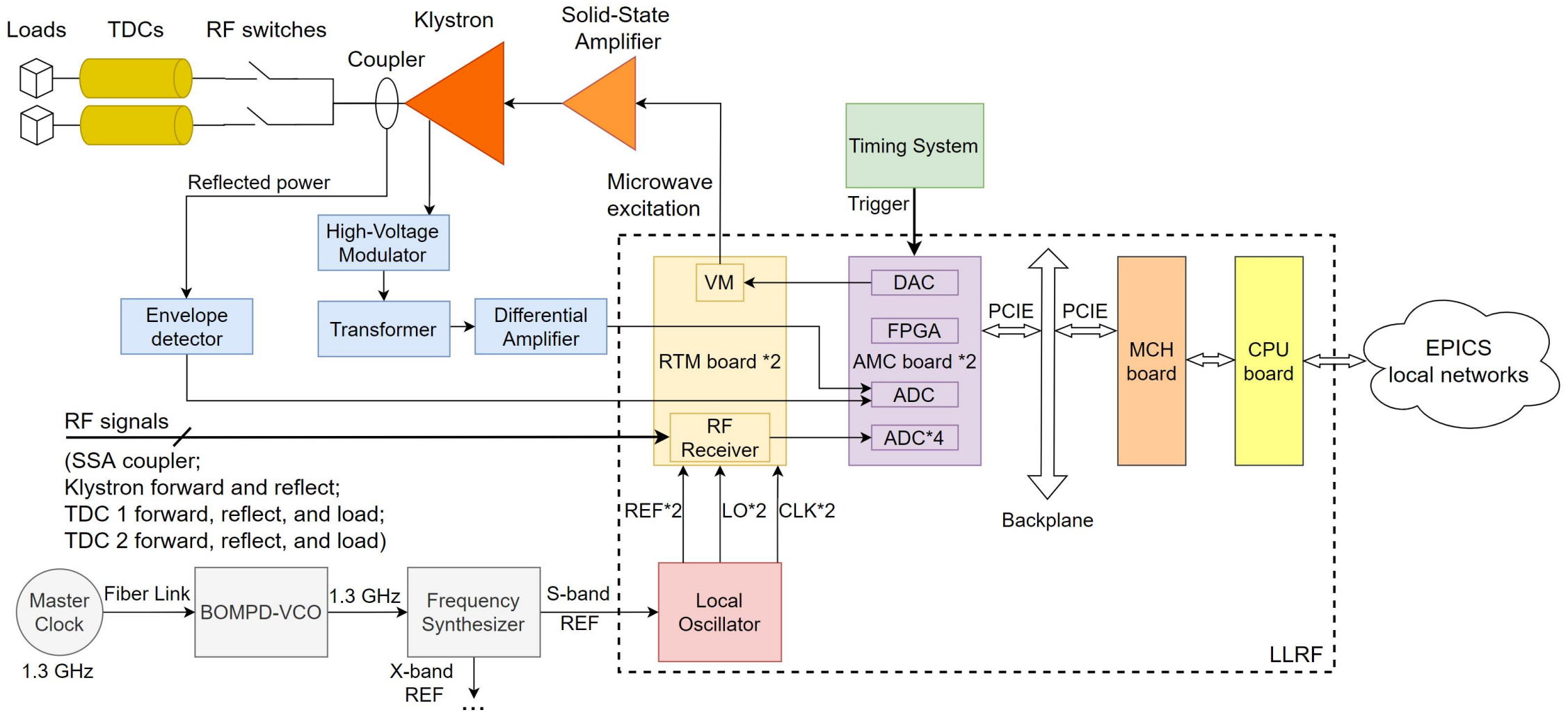

**Figure 2.** The block diagram of the LLRF system and its connected external systems.

The local oscillator converts the synchronization system's RF reference signal into two reference signals (REF), two LO signals, and two sampling clock signals (CLK) through frequency conversion. Each REF, LO, and CLK signal is connected to one RTM board. Two RTM boards mix the local oscillator clock signal with RF signals from the accelerator system and the reference signal from the synchronization system. They generate intermediate frequency (IF) signals. These IF signals are then digitized via Analog-to-Digital Conversion (ADC) and transmitted to the digital processor in two AMC boards. The RF signals include the SSA coupler's output, the klystron's forward and reflect, and two TDCs' forward, reflect, and load. We use the Field Programmable Gate Array (FPGA) as the digital processor. The software (introduced in section 3) will convert the amplitude and phase values set by operators at the microwave operating point into vectors I and Q. They will be transmitted to the FPGA through the PCIE interface and then sent to Digital-to-Analog Conversion (DAC). They are then converted to a modulated REF through a Vector Modulator (VM) to drive the SSA. After the SSA preliminarily amplifies it, it excites the klystron to work. The klystron's final output microwave is transferred through the coupler and waveguide and switched into two TDCs, which have variable polarization directions for diagnosing electron beam clusters. The reflected power of the klystron after the envelope detector is collected by a DC coupling channel of the LLRF. In addition, the sampling signal of the klystron high-voltage modulator is transformed and amplified and finally collected through another DC coupling acquisition channel. The FPGA can transmit the processed pulse microwave amplitude and phase to the CPU through the backplane and MCH PCIE interface and publish them as Process Variable (PV) to the Experimental Physics and Industrial Control System (EPICS) network for slow control and data acquisition.

Figure 3 presents the constant temperature cabinet housing the LLRF hardware. This hardware, including the LO, differential amplifier, 2U MTCA crate, SSA, and more, is arranged from top to bottom. The signal source within this cabinet is used for the LLRF calibration. At the bottom of the cabinet, the refrigeration unit plays a vital role in air cooling and heat dissipation, ensuring the system's temperature is maintained.

Figure 4 illustrates the diagram of the S-band local oscillator. We use ZA4PD-4-S+ [31] and ZA4PD-2-S+ [32] as the 1-4 and 1-2 power dividers for REF and LO signals, respectively. TQP3M9009 [33] is our selected radio-frequency amplifier due to its high linearity gain and surface-mount package. The REF is processed through an 83-frequency divider UXN14M9P [34], a triple multiplier based on PLL, and then divided by four using HMC705 [35] to generate an IF signal of 27.08 MHz. This signal enters the mixer HMC213 [36] and ultimately produces the LO. Similarly, the reference signal is divided by frequency to generate a clock CLK for ADC acquisition, ensuring a ratio of 3:13 between IF and CLK, which is used to achieve a specific IQ demodulation (will be introduced in section 3) in FPGA.

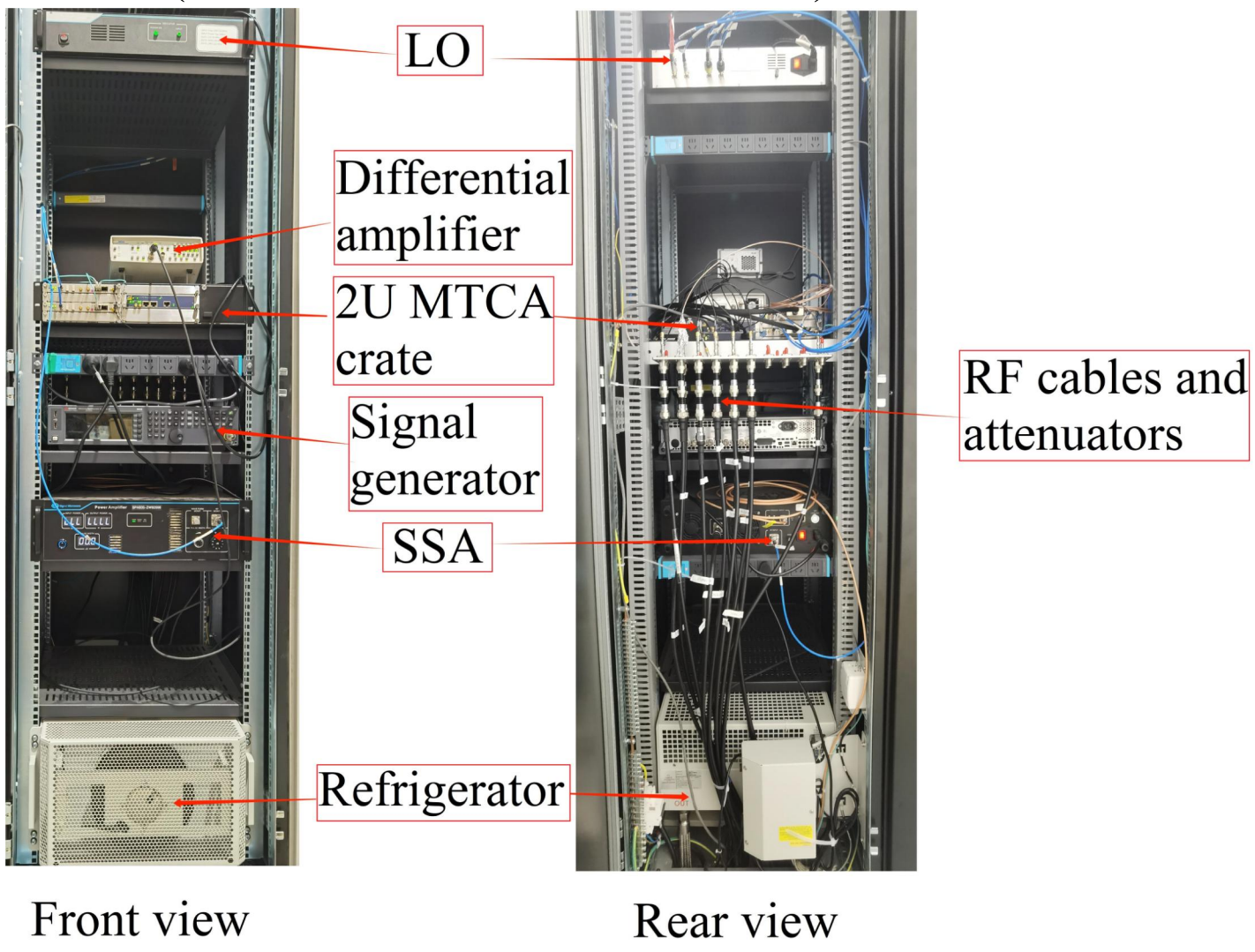

**Figure 3.** The constant temperature cabinet and its internal pieces of equipment.

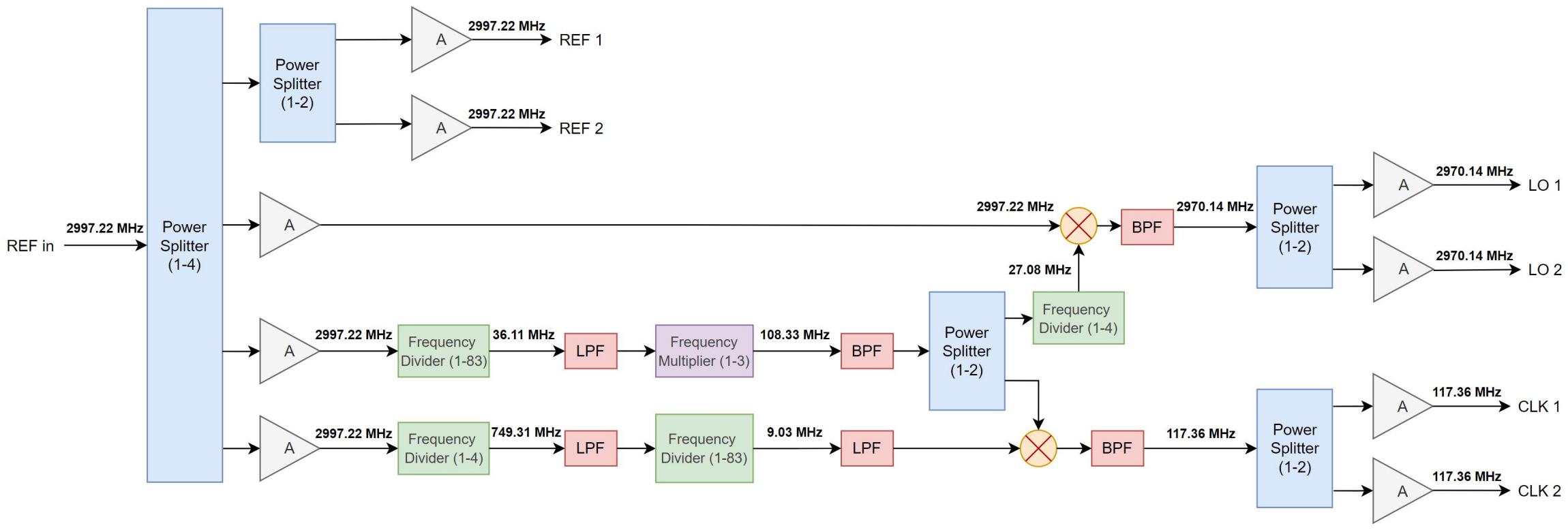

**Figure 4.** The diagram of the LO.

## 3. Firmware and software design

The FPGA firmware design implements primary algorithms. Its functions mainly include trigger selection, amplitude and phase demodulation of microwave signals, amplitude and phase stability control, and microwave excitation output for specific amplitude and phase. Figure 5 shows the design diagram of the FPGA firmware.

There are eight AC coupling inputs (AC0~AC7) and two DC (DC0~DC1) coupling input ports. Four ADCs will sample RF signals after down-conversion (by mixers), and one ADC will directly monitor the klystron's high-voltage modulator and reflected power. When the reflected power surpasses a specific threshold, the RF switch preceding the VM output will be disconnected. AC6 and AC7 ports are respectively used for monitoring the REF and VM. The clock and data recovery process is implemented by Xilinx primitives: the IBUFDS primitive converts differential clocks and data into single-ended signals. Following the IBUFDS stage, the BUFG primitive generates a global clock signal at 117.36 MHz. The single-ended data is then converted from serial to parallel format using the IDDR primitive. The basic IQ demodulation formulas are shown in equation (3.1). In our application, the frequency ratio of IF to CLK is 3:13, with n being 13. The coefficients for multiplying vector I and Q are stored in FPGA ROM.

$$
\begin{aligned}
I_k &= \frac{2}{n}\sum_{l=k-n+1}^{k} x_l \sin\left(l\Delta\varphi\right), \\
Q_k &= \frac{2}{n}\sum_{l=k-n+1}^{k} x_l \cos\left(l\Delta\varphi\right) \\
\textit{where } \Delta\varphi &= 2\pi f_{IF} / f_{CLK} = 6\pi/13,\ n = 13
\end{aligned}
\qquad (3.1)
$$

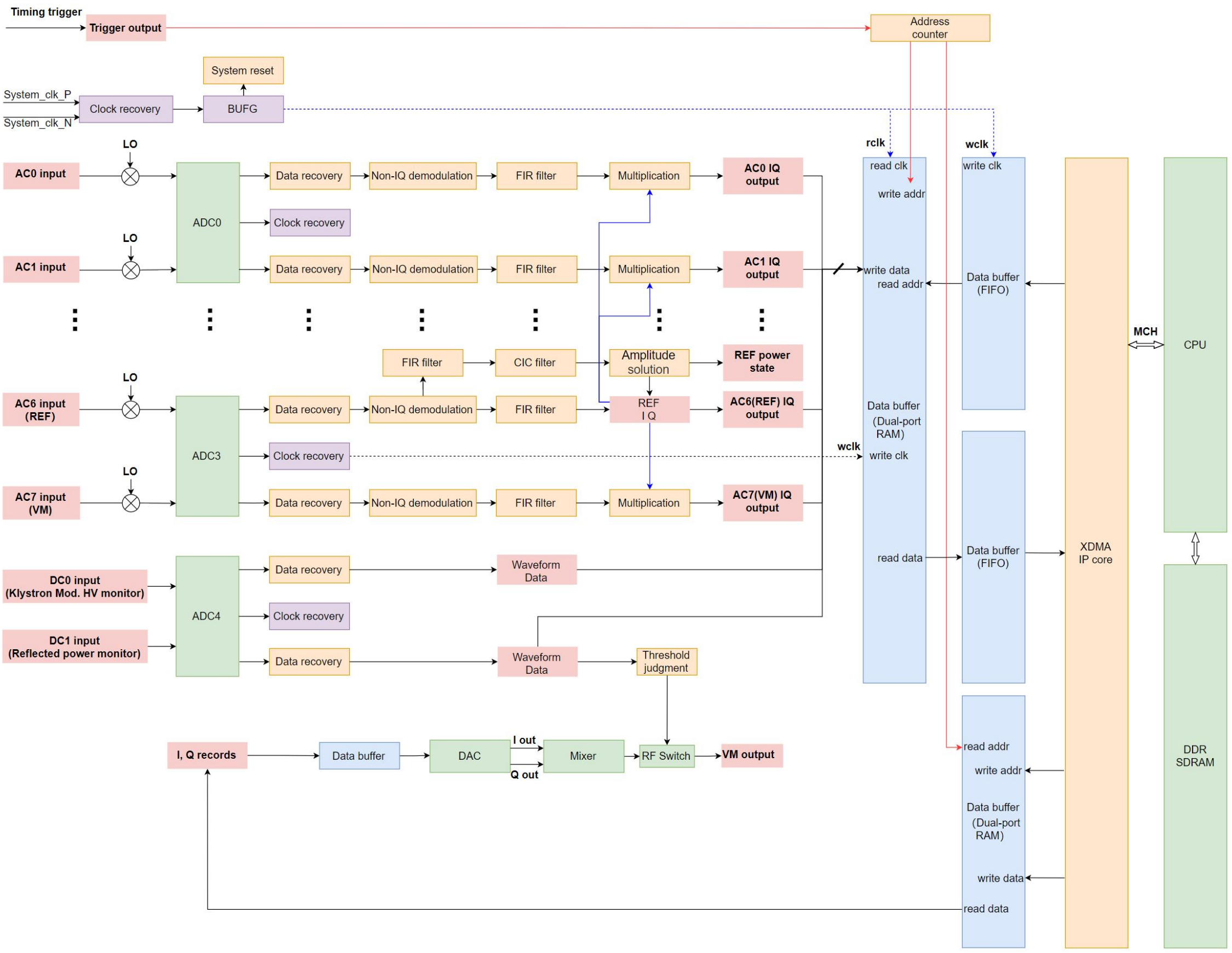

**Figure 5.** The design diagram of the FPGA firmware.

After demodulation, the signal is smoothed and buffered through a 13-order FIR filter. For port AC6, we use a CIC filter to obtain the average reference amplitude. Equation (3.2) is the general reference tracking algorithm. It eliminates the drift in the LLRF clock allocation network, which causes the measurement drift in microwave detection. We can make phase subtraction based on Euler's formula, as shown in equation (3.3) [37]. The multiplication of complex numbers is processed in firmware.

$$\varphi'_{mea} = \varphi_{mea} - \varphi_{ref} \quad (3.2)$$

$$A_{ref}e^{-j\varphi_{ref}} = A_{ref}\cos\varphi_{ref} - jA_{ref}\sin\varphi_{ref} = I_{ref} - jQ_{ref}$$

$$A_{mea}e^{j\varphi_{mea}} \times A_{ref}e^{-j\varphi_{ref}} \times \frac{1}{A_{ref}} = \frac{A_{ref}}{A_{ref}} A_{mea}e^{j(\varphi_{mea}-\varphi_{ref})} = A_{mea}e^{j(\varphi_{mea}-\varphi_{ref})} \quad (3.3)$$

where $\left(\frac{A_{ref}}{A_{ref}} \approx 1\right)$

As shown in figure 5, the PCIE clock is from the system clock (125 MHz). The rclk and wclk stand for read clock and write clock. The dual-port RAM facilitates data buffering across different clock domains, and the Xilinx XDMA is used to implement PCIE links. Upon receiving a timing trigger, we will initiate the address counter, load the amplitude and phase excitation settings configured by the user into the FPGA, and subsequently write them down to the DAC. Additionally, we will read data from each acquisition channel into memory via PCIE.

Figure 6 displays the preliminary design of the LLRF software GUI. The software design is based on EPICS architecture, using the Python language for interface design. It facilitates slow control, status presentation, and the display of microwave power and phase at each port of LLRF. Users can manipulate the LLRF and manage the turn on and off of triggers via the GUI. Users can also monitor the timing trigger status and frequency, REF signal status, and the reflected power of the klystron.

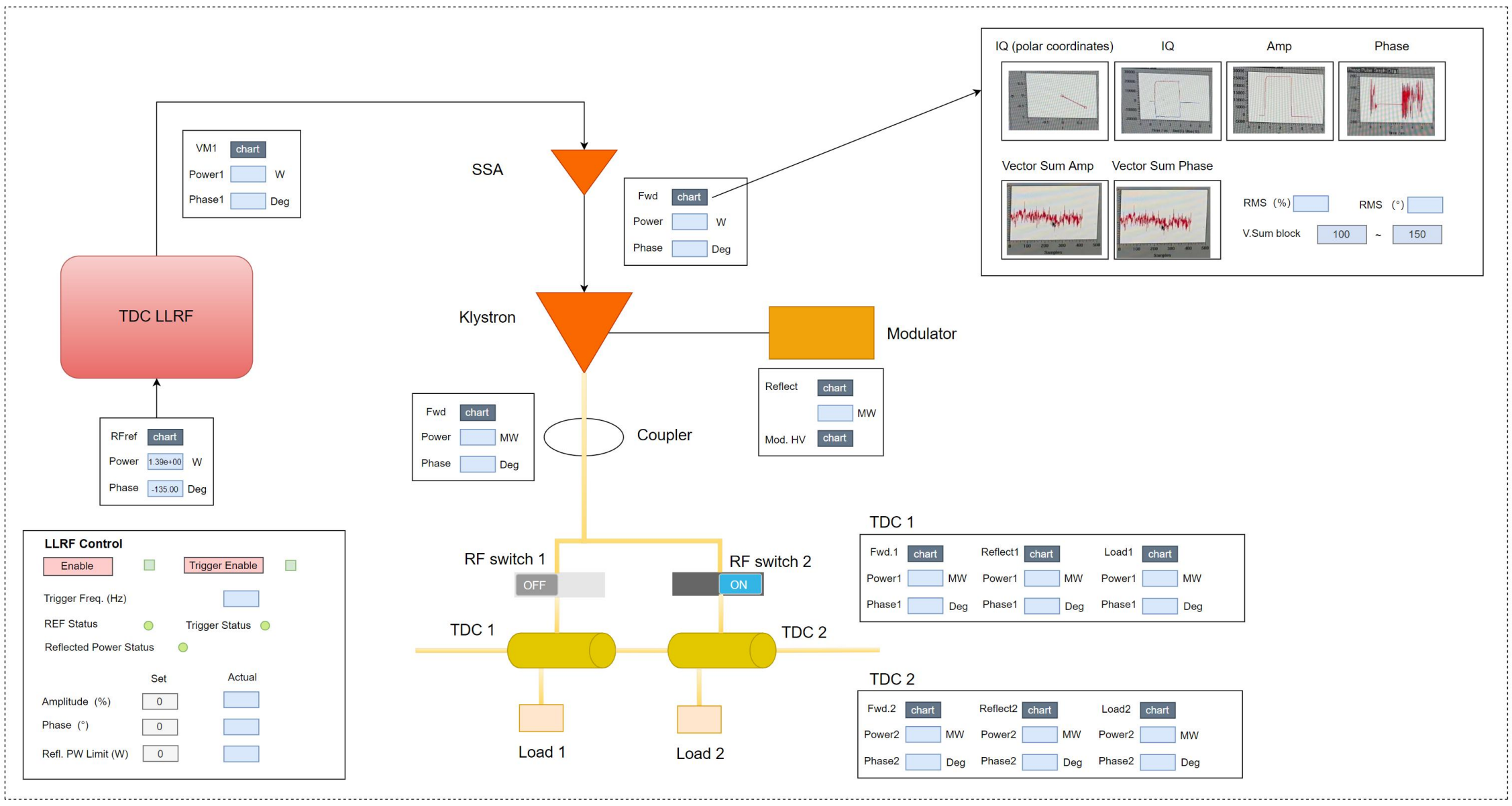

**Figure 6.** The preliminary design of the LLRF software GUI.

Furthermore, it allows for configuring the amplitude and phase of the LLRF VM output and setting a protection threshold for the klystron's reflected power. The preset values and the actual readback data must be presented on the interface. Additionally, waveforms from each acquisition channel must be displayed, encompassing waveforms in IQ polar and cartesian coordinate systems, waveforms depicting amplitude and phase, and the variation and stability (RMS) of intra-pulse integration for amplitude and phase over time.

## 4. Preliminary experiments and results

We have conducted preliminary experiments combining the synchronization system, timing system, LLRF, and SSA. Considering the filling time of the $S^3$FEL traveling-wave tube, the pulse's average amplitude and phase with a length of 380 ns (4.00 μs ~ 4.38 μs) are selected. The microwave amplitude and phase stability of the SSA are $(0.04\pm0.001)\%$ / $(0.04\pm0.001)^{\circ}$ (RMS), as depicted in figure 7. The amplitude and phase waveform with a pulse width length of 3 μs are also displayed.

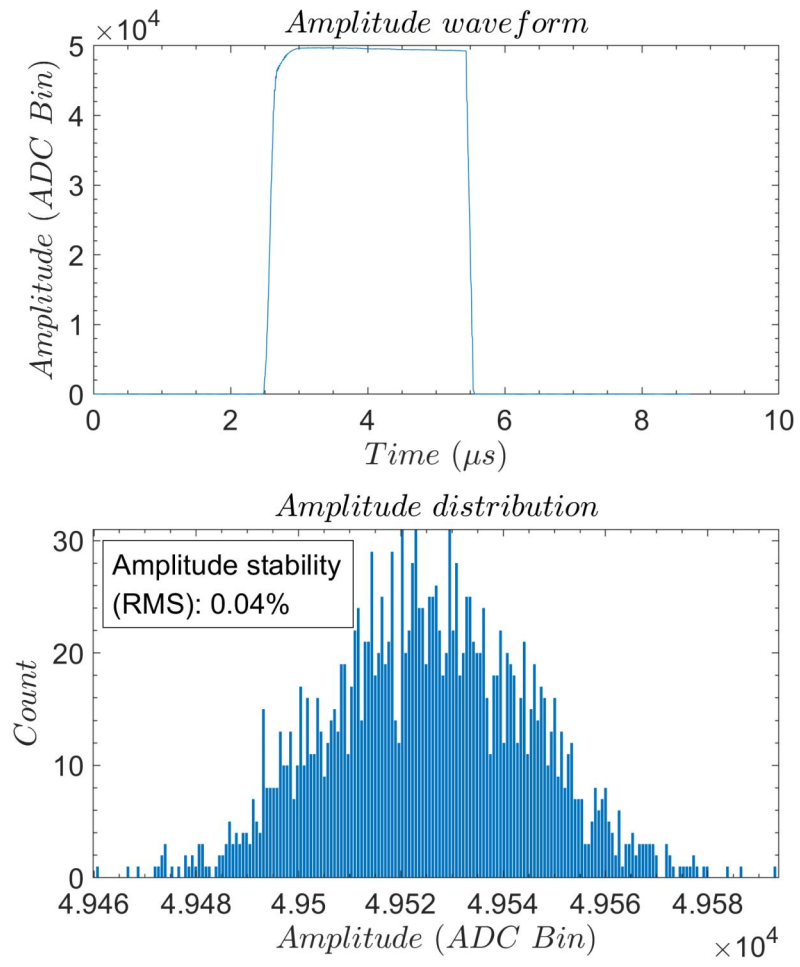

**Figure 7.** The preliminary tests of the microwave system.

Figure 8 shows the phase jitter of the REF sampled at the LLRF turning off and on the reference tracking algorithm. We place the LLRF in the chassis and open the refrigeration unit to maintain the cabinet's temperature stability within 25 $\pm$ 0.1 degrees Celsius. As observed in the graph, after turning off the reference tracking algorithm for half an hour, the phase of the LLRF acquisition shifted by 0.05 degrees. The reference tracking algorithm effectively suppresses the slow phase drift of the LLRF acquisition channel.

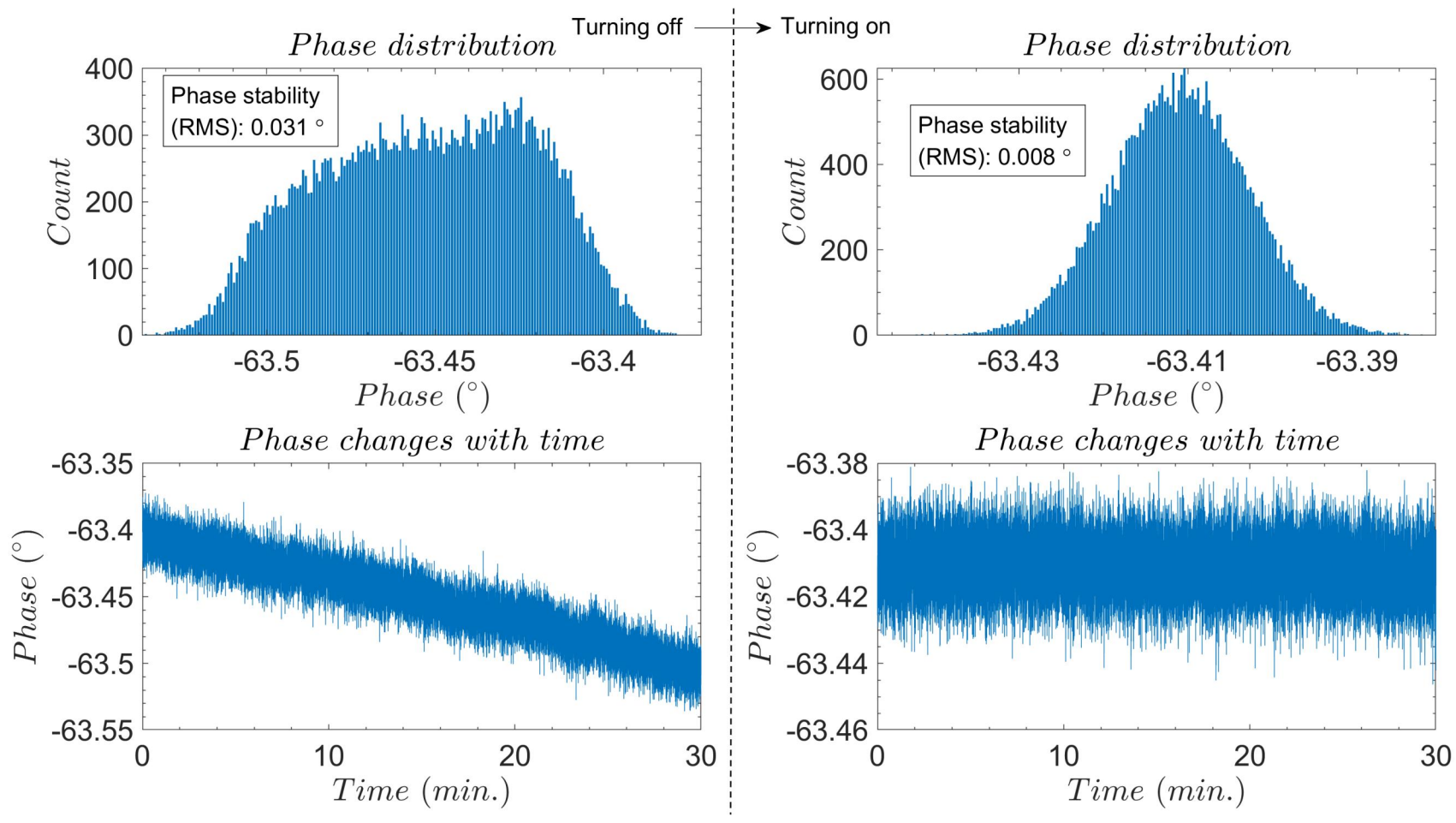

**Figure 8.** The effectiveness of proof experiments for the reference tracking algorithm.

## Acknowledgments

This work is supported by the National Natural Science Foundation of China (Grant No. 12405221), and the Shenzhen Science and Technology Program (Grant No. RCBS20221008093247072).

We would like to thank those who collaborated on the DCLS and $S^3$FEL.